\documentclass[prl,twocolumn,aps,showpacs,nofootinbib,nobibnotes,superscriptaddress,preprintnumbers]{revtex4}
\usepackage{epsfig}
\usepackage{graphics}
\usepackage{bm}
\usepackage{color}
\usepackage{dcolumn}   
\usepackage[spanish,english]{babel}
\usepackage{bm}     
\usepackage{bbm}       
\usepackage{amssymb}  
\usepackage{amsmath}
\usepackage{latexsym}
\usepackage{float}
\usepackage{ifthen}
\usepackage{caption,subfig}
\usepackage{enumerate}
\usepackage{url}
\usepackage{caption,subfig}
\usepackage{amsopn}
\usepackage{hyperref}

\usepackage{amsfonts}
\usepackage{multirow}
\usepackage{array}
\usepackage{booktabs}
\usepackage{rotating}

\usepackage{ulem}
\def\clap#1{\hbox to 0pt{\hss#1\hss}}

\def\bea{\begin{eqnarray}}
\def\eea{\end{eqnarray}}
\def\be{\begin{equation}}
\def\ee{\end{equation}}

\def\d{\mathrm{d}}

\newcommand{\Fm}{{\mathcal F}}

\newcommand{\Ft}{\tilde{F}}

\newcommand{\cgw}{c_{\rm gw}}

\newcommand{\LL}{\mathcal{L}}

\renewcommand{\geq}{\geqslant}
\renewcommand{\leq}{\leqslant}

\newcommand{\mS}{\mathcal{S}}
\newcommand{\mK}{\mathcal{K}}

\begin{document}

\title{Non-trivial gravitational waves and structure formation \\phenomenology from dark energy}

\date{\today,~ $ $}

\author{Jose Beltr\'an Jim\'enez} 
\email{jose.beltran@uam.es}
\affiliation{Instituto de F\'isica Te\'orica UAM-CSIC, Universidad Aut\'onoma de Madrid, Cantoblanco, Madrid, 28049, Spain.}
\affiliation{Departamento de F\'isica Fundamental, Universidad de Salamanca, E-37008 Salamanca, Spain.}

\author{Lavinia Heisenberg} \email{lavinia.heisenberg@eth-its.ethz.ch}
\affiliation{Institute for Theoretical Studies, ETH Zurich, 
 Clausiusstrasse 47, 8092 Zurich, Switzerland}

\date{\today}

\begin{abstract}
The detection of the GW170817/GRB170817A event improved the constraints on the propagation speed of gravitational waves, thus placing possible variations caused by dark energy under restraint. For models based on scalar fields belonging to the family of Horndeski Lagrangians, non-minimal derivative couplings are now severely constrained, entailing a substantially limited phenomenology. In this work we want to stress that there is still a plethora of dark energy models that get around this obstacle while still providing interesting phenomenologies able to distinguish them from the standard cosmology. We focus on a class involving vector fields as a proxy, but our discussion is extensible to a broader class of models. In particular, we show the possibility of having a non-minimal derivative coupling giving a non-trivial effect on scalar modes without affecting gravitational waves and the possibility of having a second tensor mode that can oscillate into gravitational waves. We also present a novel class of configurations breaking rotational invariance but with an energy-momentum tensor that is isotropic on-shell. This peculiar feature makes the scalar and vector sectors of the perturbations mix so that, even in a perfectly isotropic background cosmology, preferred direction effects can appear in the perturbations.  We also comment on models that give rise to isotropic solutions when averaging over rapid oscillations of the vector fields. The explored models are classified according to distinctive field configurations that provide inequivalent realisations of the Cosmological Principle.

\end{abstract}


\maketitle

\section{Introduction}

General Relativity (GR) stands out as the most compelling contender to describe the gravitational interaction in a wide range of scales, from sub-milimiter to Solar System scales \cite{Will:2014kxa}. Most experiments designed to test GR have focused on its non-radiative sector, aiming to finding deviations from Newton's law, the equivalence principle or, in general, in any of the Parameterised Post-Newtonian (PPN) parameters. Until recently, the radiative sector, namely Gravitational Waves (GWs), has remained largely less constrained from direct means, partly due to its elusive character. However, this difficulty has not prevented to have constraints from indirect probes, being the variation in the period of binary pulsar systems the most clear indirect evidence for the existence and properties of the GWs (see e.g. \cite{Wex:2014nva}). In fact, these measurements already allowed to infer that the gravitational radiation is predominantly quadrupolar with an amplitude in perfect agreement with the GR predictions, implying that GWs have spin 2,  and its propagation speed $\cgw$ could only differ from the speed of light $c$ at the $10^{-2} - 10^{-3}$ level, with the corresponding bounds for modified gravity theories \cite{Yagi:2013qpa,Jimenez:2015bwa}. The speed of (subluminal) GWs was also constrained from the absence of Cherenkov radiation for cosmic rays to the $10^{-15}$ level ($10^{-19}$ for cosmic rays of extra-galactic origin) \cite{Moore:2001bv}. These Cherenkov radiation constraints were already used in \cite{Kimura:2011qn} to set bounds within the Horndeski scalar-tensor theories relevant for dark energy models.

The situation improved dramatically after the first direct detections of GWs by the LIGO team, clearly confirming their physical reality and proving the feasibility of GWs astronomy. In August 2017, the VIRGO team joined the LIGO network, increasing the sensitivity to the polarisation of the GWs that provided a direct confirmation of its spin-2 nature\footnote{The analysis was performed by assuming pure tensor, pure vector or pure scalar GWs with the case of pure tensor being strongly favoured. This however does not exclude the existence of additional polarisations that could be subdominant (binary pulsars also indicate that a possible radiation in scalar or vector modes must be strongly suppressed with respect to the quadrupolar emission in GWs). It has also been argued in \cite{Allen:2018khw} that a pure vector could, in principle, fit the LIGO/VIRGO signal, although a specific polarisation evolution needs to be assumed and it is unclear if it can be put into effect in a realistic theoretical framework.} of GWs after only one detection \cite{Abbott:2017oio}.  Another major (and perhaps the most outstanding so far) discovery in GWs astronomy was the GW170817 event corresponding to a merger of two neutron stars \cite{TheLIGOScientific:2017qsa}. What made this detection even more exciting was the possibility of identifying and observing the same event in the electromagnetic channel as the GRB170817A signal. Among many other outstanding consequences, this multi-messenger detection provided a direct constraint on the propagation speed of GWs to be $\vert \cgw/c-1\vert\lesssim10^{-15}$ in agreement with the previous indirect bounds and confirming once again the predictions of GR. This constraint is many orders of magnitude tighter than the ones obtained from binary pulsars and it is at the same level as the ones drawn from the absence of Cherenkov radiation, although it directly applies to much lower frequencies. 

The constraints from GWs astronomy (even if only a few events are available so far) together with the indirect observations of binary pulsars and the classical tests of GR on Solar System scales do not show any deviations from the GR predictions and, thus, theories of modified gravity are tightly constrained in the infrared nowadays \cite{Berti:2015itd}. This is specially important for models of dark energy based on modified gravity theories because it seems harder and harder to have models that can provide accelerated expansion on cosmological scales while being compatible with all local gravity tests. A way out to this dichotomy was the existence of screening mechanisms that could allow to hide the dark energy effects on small scales (see e.g. \cite{Joyce:2014kja} for a comprehensive review on this subject). As a paradigmatic class of modified gravity theories for dark energy we can consider scalar-tensor theories and, in particular, the extensively studied family of Horndeski Lagrangians \cite{Horndeski:1974wa} and some of its extensions \cite{beyondH}. These theories are characterised by the presence of second order derivative self-interactions of the scalar field that, in turn, require derivative non-minimal couplings to the gravity sector in order to ensure the absence of additional ghost-like propagating modes. These non-minimal couplings give rise to a very rich phenomenology for dark energy models (which is nicely captured in the effective field theory of dark energy \cite{EFTDE}), but the very same operators that drive the interesting phenomenology for structure formation are responsible for a variation of the propagation speed of GWs. The main aim of the present work is to put forward a class of dark energy models realising one of the following features
\begin{itemize}
\item Non-minimal derivative couplings with effects on the scalar perturbations without affecting the GWs sector.
\item Non-trivial predictions for GWs astronomy without immediately conflicting with $\cgw=1$.
\end{itemize}
The first condition contrasts with the findings for dark energy models based on the Horndeski Lagrangians discussed above, where non-minimal derivative couplings of the scalar field gives rise to a modification of $\cgw$ in the presence of a time-dependent background for the scalar field and, therefore, are subject to the constraint imposed by the GW170817/GRB170817A observation \cite{Ezquiaga:2017ekz}. On the other hand, the second condition shows that there is still room to probe dark energy models with GWs astronomy in a non-trivial way without being in tension with $\cgw=1$. In particular, the models that we will discuss give rise to a possible oscillation of GWs into additional tensor modes. Throughout this work we will use theories with vector fields as proxies to show the different possibilities, but the same can be applied to other models, as we will discuss in the last section. These different possibilities will be fundamentally characterised by the crucial property of providing inequivalent realisations of the Cosmological Principle, i.e., the homogeneity and isotropy of the universe will be achieved from different symmetries owed to different field configurations. 

Before proceeding to the main discussion, let us fix some notation. Given a set of vector fields $A^a{}_\mu$ with $a$ denoting some internal index and $\mu$ a Lorentz/spacetime index, we will define the field strengths as $F^a_{\mu\nu}=\partial_\mu A^a_{\nu}-\partial_\nu A^a_{\mu}$ and $\Ft^{a\mu\nu}=\frac12\epsilon^{\mu\nu\rho\sigma}F^a_{\rho\sigma}$ the corresponding dual tensor. From these objects we can define the electric and magnetic components as $E_i=F_{0i}$ and $B_i=\Ft_{0i}=\frac12 \epsilon_{ijk}F_{jk}$. In the case of gauge fields, the field strength will be denoted by $\mathcal{F}^a_{\mu\nu}=F^a_{\mu\nu}+gf^{abc}A^b{}_\mu A^c{}_\nu$, with $g$ the coupling constant and $f^{abc}$ the structure constants of the corresponding gauge group. We will also use the symmetrised covariant derivative given by $S^a_{\mu\nu}=\nabla_\mu A^a_{\nu}+\nabla_\nu A^a_{\mu}$. The norm of the vector fields will be denoted by $Y=A^a{}_\mu A^{a\mu}$. Furthermore, we will consider cosmological models described by the FLRW metric $\d s^2=\d t^2-a^2(t)\d \vec{x}^2$ with $H=\dot{a}/a$ the corresponding Hubble expansion rate. We will use both a bar and a subscript $0$ to denote the background value of some quantity.

\section{Dark Energy in multi-Proca and Yang-Mills theories}
As explained in the introduction, we will use theories featuring vector fields to illustrate the dark energy models complying with our requirements. More specifically, we will consider theories characterised by the presence of a set of $N$ vector fields $A^a{}_\mu$, with $a=1,\cdots N$. For simplicity we will only consider $N=3$ and, in order to ensure the existence of homogeneous and isotropic solutions, we will assume an internal global $SO(3)$ symmetry. Thus, our Lagrangians will be built out of $SO(3)$ and Lorentz scalars involving the vector fields and up to their first derivatives, i.e., $A^a_{\mu}$, $F^a_{\mu\nu}$ and $S^a_{\mu\nu}$. For a detailed spelled out of the possible terms in the Lagrangian we refer to \cite{Allys:2016kbq,Jimenez:2016upj}. In most of this work we will however focus on the simplest term given by an arbitrary function $\mK$ involving only the vector fields $A^a_\mu$ and their field strengths $F^a_{\mu\nu}$, unless otherwise stated. The global symmetry can also be promoted to a gauge symmetry (that can then be identified with an $SU(2)$ gauge symmetry) in which case we will be dealing with Yang-Mills theories. In that case, the theories can be built in terms of the $SU(2)$ scalars involving $\Fm^a_{\mu\nu}$.  Although we will restrict ourselves to these groups, it is worth mentioning that any larger group within which these can be embedded will also work. For instance, we could consider GUT groups such as $SU(5)$ or $SO(10)$ that contain our desired groups as subgroups. Of course, the precise low energy phenomenology will depend on the specific symmetry breaking pattern, but what concerns us here will be to have an internal symmetry that can be traded by the breaking of spacetime symmetries so that we can eventually have a residual $ISO(3)$ symmetry complying with the Cosmological Principle. Within this set-up, we can now consider different field configurations to achieve a homogeneous and isotropic cosmological background as we discuss in detail in the following.

\subsection{Pure temporal configuration}
We will start with the simplest homogeneous and isotropic configuration for the fields given by
\be
A^a{}_\mu=\phi^a(t)\delta^0{}_\mu,
\ee
with $\phi^a(t)$ arbitrary functions of time. This field configuration actually corresponds to a version of several single Proca fields and, in fact, no internal symmetry is required to have a cosmological background. Thus, it suffices to analyse the single field case to pinpoint the relevant phenomenology so that we will drop the internal index in the following. The interest of this configuration is the existence of a non-minimal derivative interaction for the vector field of the form\footnote{The multi-vector case simply amounts to adding internal indices.}
\be
\LL\supset G_6(Y)L^{\mu\nu\alpha\beta} F_{\mu\nu}F_{\alpha\beta}
-G'_{6}(Y) \tilde{F}^{\alpha\beta}\tilde{F}^{\mu\nu}S_{\alpha\mu} S_{\beta\nu} ,
\label{L6}
\ee
with $L^{\mu\nu\alpha\beta}=\frac14 \epsilon^{\mu\nu\rho\sigma}\epsilon^{\alpha\beta\gamma\delta} R_{\rho\sigma\gamma\delta}$ the double dual Riemann tensor, $G_6$ an arbitrary scalar function depending on $Y$ and a prime standing for derivative w.r.t. its argument. The case $G'_6=0$ corresponds to the vector-tensor interaction already found by Horndeski\footnote{Some cosmological consequences and stability properties were explored in \cite{Barrow:2012ay, Jimenez:2013qsa}.} \cite{Horndeski:1976gi}. This non-minimal coupling has no analogous in the Horndeski scalar-tensor Lagrangians and, thus, it can give rise to novel phenomenological effects. Since the background configuration has an identically vanishing field strength $\bar{F}_{\mu\nu}=0$, the contribution to the quadratic Lagrangian for the perturbations is simply
\begin{align}
\LL^{(2)}\supset&\Big[G_6(Y)R^{\mu\nu\alpha\beta} 
+\frac{G'_{6}(Y)}{2}S^{\mu\alpha} S^{\nu\beta}\Big]_0\delta\Ft_{\mu\nu}\delta\Ft_{\alpha\beta}\nonumber\\
=&-\frac{4H^2}{a^2}\left(\bar{G}_6+2\phi^2 \bar{G}'_6\right)\delta \vec{E}^2\nonumber\\
&+\frac{4}{a^4}\left[(H^2+\dot{H})\bar{G}_6+2H\phi\dot{\phi}\bar{G}'_6\right]\delta \vec{B}^2 .
\label{L6pert}
\end{align}
This expression clearly shows that there are no effects on the GWs sector at this order. In fact, all metric perturbations are trivially absent from this term and only the vector field perturbations contribute, which however have a non-trivial impact on the perturbations. Obviously, it gives rise to propagating vector perturbations (possibly together with other terms in the full Lagrangian as e.g. a standard Maxwell kinetic term). Concerning the scalar sector, it gives a non-trivial contribution to the effective gravitational Newton's constant. The crucial point to understand how these effects are generated is to remember that the temporal component of the vector is an auxiliary field in these theories. Thus, although metric perturbations do not contribute in (\ref{L6pert}), the perturbation of $A_0$ does and, after integrating it out, it will give rise to a modification of the scalar sector. We should say however that a genuine contribution to the effective Newton's constant from the Horndeski interaction requires the presence of a term like $G_3(A^2)\nabla_\mu A^\mu$ that will give rise to a braiding similar to the KGB models for scalar fields \cite{Deffayet:2010qz}, but with additional operators contributing to it. Let us also mention that this term is also necessary for having a dynamical background evolution for the vector field. In order to illustrate and clarify these points, let us consider the Lagrangian\footnote{We could also add a term $G(Y)\tilde{F}^{\mu\alpha}\tilde{F}^\nu{}_{\alpha} S_{\mu\nu}$ which does not modify the background evolution but can affect the perturbations. This term does not add anything crucially new to our discussion and, nevertheless, we want to focus on the non-minimal derivative coupling. Let us also mention that going to the multi-vector extension allows to introduce additional terms linear in $S^a_{\mu\nu}$ that are not present in the single field case, as e.g. $\epsilon^{\alpha\beta\gamma\delta}\tilde{F}^a_{\alpha\lambda} S^{b\lambda}{}_\beta A^a_\gamma A^b_\delta$ or $S^{a\mu\nu} A_\mu^b A_\nu^d A_\alpha^c A^{e \alpha} \delta_{de} \epsilon_{abc}$. All these terms could give rise to additional interesting features, but they are not crucial for our purpose here.} 
\begin{align}
\LL=&\mK(Y,\cdots)+G_3(Y)\nabla_\mu A^\mu\nonumber\\
&+G_6(Y)L^{\mu\nu\alpha\beta} F_{\mu\nu}F_{\alpha\beta}
-G'_{6}(Y) \tilde{F}^{\alpha\beta}\tilde{F}^{\mu\nu}S_{\alpha\mu} S_{\beta\nu}\,,
\label{eq:action1}
\end{align}
where the dots stand for terms involving $F_{\mu\nu}$ that will not contribute to the background equations of motion, although they will affect the perturbations. For the case of one single vector field (as we are considering here), there are only two possibilities, namely $F_{\mu\nu}F^{\mu\nu}$ and $A^\mu A^\nu F_{\mu\alpha} F^\alpha{}_\nu$, but the multi-vector case allows for more terms some of which can also contribute to the background equations very much like $Y$. 

One might worry that the non-minimal coupling to the double dual Riemann tensor will introduce modifications to $\cgw$ whenever the field strength has a non-vanishing profile, as it would be expected on sub-Hubble scales, and this could conflict with the tight constraint on $\cgw$. We however see this as good news since it gives the possibility to test this coupling with GWs. With only one measurement it is difficult to say anything definitive, since the signal might have travelled without encountering any relevant vector field profile. However, more multimessenger observations will put interesting constraints on backgrounds with a non-vanishing profile of the vector field strength. Notice however that it is crucial to have a background vector field generating a non-trivial electric and/or magnetic component.

The background vector field equations of motion for a purely temporal background derived from (\ref{eq:action1}) reduce to
\be
\phi\Big(3\phi H G'_{3} + \mK_Y\Big)=0\,,
\ee
where we see that we need $G'_3\neq 0$ in order to have an evolving $\phi(t)$. By solving the non-trivial branch ($\phi\neq 0$) of the above equation, we obtain $\phi=\phi(H)$, which can then be inserted in the Friedmann equation, modifying that way the way in which matter fields affect the expansion of the universe\footnote{Interestingly, these theories can be obtained in quadratic gravity theories formulated in Weyl \cite{Jimenez:2014rna} and general vector distorted geometries \cite{Jimenez:2015fva}. For some cosmological consequences and applications to inflation and dark energy models see also \cite{Jimenez:2016opp}.}  \cite{DeFelice:2016yws}. This is a general result whenever there are auxiliary fields in the gravitational sector (see e.g. \cite{auxiliary} for some other realisations).  

Let us begin by arguing why it is not possible to obtain any anomalous slip parameter\footnote{The slip parameter can be defined as the difference between the Newtonian and the lensing potentials or, equivalently, as the ratio of the two gravitational potentials in the Newtonian gauge. The precise definition is not important for us.}. A departure from the GR value of the slip parameter can be traced to the presence of anisotropic stresses, since the off-diagonal spatial gravitational equations lead to $\Phi-\Psi\propto \sigma$, being $\Phi$ and $\Psi$ the gravitational potentials and $\sigma$ the scalar potential of the anisotropic stress. Thus, it is easy to convince oneself that we need to have an energy-momentum tensor such that $T_{ij}$ is not proportional to $\delta_{ij}$. Let us then see what quantities could contribute to $T_{ij}$ in the considered background. Since $F_{ij}$ is antisymmetric and its background value vanishes, it is obvious that it can not contribute to the scalar perturbation of $T_{ij}$ at first order. Quantities involving only the vector without derivatives will contribute as $A_i A_j$ so, again, no first order contributions are possible. Thus, we are left with contributions involving $S_{\mu\nu}$. If we want to keep $\cgw=1$, then only the aforementioned term $G_3\nabla_\mu A^\mu$ can be included. However, the variation of this term w.r.t. the metric gives 
\be
\delta\big(\sqrt{-g} G_3\nabla_\mu A^\mu\big)=\delta G_3\partial_\mu(\sqrt{-g}A^\mu)-\partial_\mu G_3\delta(\sqrt{-g} A^\mu)
\ee
up to integrations by parts. We thus see that it is not possible to obtain a term in the energy-momentum tensor of the form $\partial_i\delta A_j$. Having explored the different possible contributions to the anisotropic stress and obtaining that none of them can actually contribute in the temporal background, we then conclude that no anomalous slip parameter can arise by maintaining $\cgw=1$. This is along the lines of the conclusions reached in \cite{Saltas:2014dha}.

Let us now turn to non-trivial effects. Even though the non-minimal coupling does not modify $\cgw$ with the pure temporal configuration of the vector field, it can affect the formation of large scale structures through a modification of the effective Newton's constant $G_{{\rm eff}}$, which, in the deep quasi-static approximation, takes the form \cite{DeFelice:2016yws}
\be
\frac{G_{\rm eff}}{G}=\frac{(\alpha_1G_3'+\alpha_2 G_3'')(G_6+2\phi^2G_6')+\mathcal{F}_1}{(\beta_1G_3'+\beta_2 G_3'')(G_6+2\phi^2G_6')+\mathcal{F}_2}\,,
\ee
where $G$ is the usual Newton's constant, $\alpha_{1,2}$ and $\beta_{1,2}$ are functions of $\{\phi, \dot{\phi},H,\dot{H}\}$ and $\mathcal{F}_{1,2}$ are functions of those same variables and also depend on $\mK$ and its derivatives. The specific form of these functions is not relevant for us here, but we have explicitly spelled out the dependence on $G_3$ and $G_6$ that confirms our statement above, namely, that it is crucial to have $G_3$ for the non-minimal coupling controlled by $G_6$ to affect the scalar perturbations. Since this is also the condition to have an evolving field background $\phi(t)$, it is in fact the expected case in the most interesting cosmologies. In \cite{Amendola:2017orw} it is shown that the effective Newton's constant within these theories is generically larger than $G$, implying an enhancement in the clustering of dark matter\footnote{The general result reported in \cite{Amendola:2017orw} that $G_{\rm eff}\geq G$ applies to general background solutions that are either asymptotically attracted to a de Sitter solution or have $w_{\rm DE}\leq-1$. The latter condition implies that is necessary to have non-phantom behaviour in order to have $G_{\rm eff}<G$. In the former case, having a background solution connecting an early phase with $G_{\rm eff}< G$ and the de Sitter attractor necessary implies crossing a point where $G_{\rm eff}$ diverges (although this divergence could occur in the future). However, more general accelerating solutions not necessarily ending in a de Sitter attractor exist for which we could have $G_{\rm eff}< G$ without ever encountering said divergence.}. It is important to emphasise however that the contribution from $G_6$ allows to introduce an additional scale to which the background evolution is oblivious, but the perturbations are sensitive to. This feature is characteristic of these vector field theories and has no analogue in the scalar field theories belonging to the Horndeski family. 

Although the effective Newton's constant is generically larger than $G$, let us comment on a possible extension where $G_{\rm eff}$ can be lowered based on a coupling of dark matter to an effective metric that depends on the vector field\footnote{A similar construction was employed in \cite{BeltranJimenez:2013fca} to develop a symmetron screening mechanism for a vector field. See also \cite{BeltranJimenez:2018tfy} for theories with a vector coupled to the energy-momentum tensor.}. The simplest of such couplings is to $\tilde{g}_{\mu\nu}=g_{\mu\nu}+A_\mu A_\nu$ so that, at linear order and around a purely temporal background, a coupling of the form $\phi(t)\delta A_0\delta\rho_{\rm DM}$ is generated, which is analogous to the usual coupling of a vector field to charged particles. Since the {\it charges} in this case are all alike and given by the mass,  this extra coupling will give rise to a repulsion in the dark matter that will result in a weakening of the structure formation. This mechanism is somewhat similar to some models of self-interacting dark matter (see e.g. \cite{DMselfinteractions}). 

We will finalise our discussion on the temporal configuration by noticing that, besides the effects on the scalar perturbations discussed thus far, it is important to keep in mind that distinctive effects will also arise in the vector perturbations and that the considered non-minimal coupling will again have an impact in them \cite{DeFelice:2016yws}.

\subsection{Triad configuration}
After discussing the purely temporal configuration, we will consider the so-called {\it triad configuration} given by 
\be
A^a{}_\mu{}=A(t)\delta^a{}_\mu
\label{triadconf}
\ee
that is still compatible with a FLRW background and represents a genuine cosmological solution for the multi-vector models that is not possible in the single field case. This configuration amounts to having three vector fields pointing along orthogonal directions that can then be identified with the three spatial axes. Thus, it breaks both the internal $SO(3)$ symmetry and the external spatial rotations, but it leaves a linear combination of them unbroken, which is then responsible for the rotational symmetry of the background. This configuration is the natural one to have cosmological solutions for non-abelian Yang-Mills theories \cite{Cervero:1978db,Galtsov:1991un,Darian:1996mb}. The simplest case of $SU(2)$ has been used for inflationary models supported by interactions of dimension higher than 4 in gauge-flation \cite{Maleknejad:2011jw} (see \cite{Maleknejad:2012fw} for a nice review) or by Horndeski-like non-minimal couplings \cite{Davydov:2015epx} (although these come at the price of tensor instabilities \cite{BeltranJimenez:2017cbn}). Needless to say that these configurations used to develop inflationary solutions can also be employed to construct dark energy models. Although there have been some proposals to describe dark energy purely in terms of Yang-Mills fields\footnote{There are also models with accelerated expansion where the gauge fields are {\it assisted} either by other fields as in e.g. \cite{Rinaldi:2015iza} or non-minimal couplings as in e.g. \cite{Bamba:2008xa}. Notice however that the latter case suffers from ghost instabilities, unless the non-minimal coupling is of the Horndeski form to the double dual Riemann tensor as in \cite{Davydov:2015epx,BeltranJimenez:2017cbn}.} with the triad configuration \cite{Zhao:2005bu,Galtsov:2008wkj,Mehrabi:2015lfa}, this has remained largely less explored than models based on scalar fields. 

The triad configuration (\ref{triadconf}) can also be used in theories without the non-abelian gauge symmetries discussed above, provided there is still a global internal $SO(3)$ symmetry allowing for a residual rotational invariance. This was already explored as a dark energy model based on a set of vector fields with a certain potential in \cite{ArmendarizPicon:2004pm} (see also \cite{Wei:2006tn}). The triad configuration was also used for models of inflation supported by massive vector fields \cite{Golovnev:2008cf}. Another class of models where this configuration is relevant is provided by the theories that extend the generalised Proca interactions \cite{Heisenberg:2014rta,Allys:2015sht,Jimenez:2016isa} (or its extensions \cite{Heisenberg:2016eld}) to the case of multiple vector fields \cite{Allys:2016kbq,Jimenez:2016upj}. The possibility of having new interesting cosmological scenarios with the triad configuration in these theories was discussed in \cite{Jimenez:2016upj}. In \cite{Rodriguez:2017wkg} some of these novel interactions have been used to show the possibility of having dark energy solutions. However, these models use non-minimal couplings that modify the propagation speed of GWs in this field configuration and, thus, their observational viability is jeopardised.

For the triad configuration (\ref{triadconf}), the non-minimal coupling to the double dual Riemann tensor discussed in the previous section gives rise to a modification of the GWs propagation speed already at the linear order and, as a consequence, it will be tightly constrained by the GW170817/GRB170817A event. Thus, we will disregard it for the triad configuration. This however does not mean that there will be a trivial effect on the propagation of GWs and, in fact, these models can actually be probed by GWs astronomy. The reason precisely roots in the symmetry breaking pattern of these models where the unbroken diagonal $SO(3)$ symmetry consists of a linear combination of the internal and the external rotations, what allows for a second tensor mode associated to the vector fields. In this case, performing the usual helicity decomposition of the perturbations (now in terms of the irreducible representations of the unbroken diagonal $SO(3)$ symmetry of the background) leads to two tensor modes, namely: the usual metric perturbation $h_{ij}=\delta_T g_{ij}/a^2$ and, in addition, the tensor mode built as $t_{ij}=\delta^a_{i} \delta_T A^a{}_j$, where $\delta_T$ stands for the transverse traceless perturbation. Notice that the perturbations of the vector fields arrange into the tensor mode $t_{ij}$  thanks to the background field $\bar{A}^a{}_i=A(t) \delta^a_i$ that allows to identify the spacetime and internal indices. The interesting feature of these models is that these two tensor modes mix in a non-trivial way and can give rise to an oscillation of GWs into the second tensor mode,  among other interesting effects. Thus, this phenomenon will give a signature of these models of dark energy that can be probed with GWs without being immediately ruled out by the tight constraint on the speed of GWs. This type of oscillations for the case of Yang-Mills fields has been explored in \cite{GWoscillationsGF}. The phenomenon of GWs oscillations also occurs in theories of massive bi-gravity \cite{GWoscillationsbiG}. However, the small value of the required mass makes such oscillations very small and the forecasted effect very difficult to detect.

The important point we want to make here is that all the dark energy models based on the triad configuration will affect in a {\it safe} way the propagation of GWs. Since we do not invoke non-minimal couplings at all, the propagation speed of GWs will be determined by the usual Einstein-Hilbert term and, thus, all these models will give rise to $\cgw=1$. However, despite being minimally coupled to gravity, the presence of the non-trivial triad configuration that breaks the spacetime and internal rotations to the diagonal component leads to a mixing of both tensor modes. For instance, a dependence on $Y$ of the Lagrangian will lead to contributions to  the quadratic action for the perturbations as
\be
\LL^{(2)}\supset \bar{A}^a_\mu\delta A^a{}_\nu\delta g^{\mu\nu}\supset A(t)\delta A_{ij}\delta g^{ij}
\ee
where $\delta A_{ij}=\delta^a_i \delta A^a{}_j$. We thus clearly see how the tensor mode associated to the vector fields mixes with the GWs through the simple $Y$-dependence of the Lagrangian. If we look at contributions arising from a kinetic dependence on $F^a{}_{\mu\nu} F^{a\mu\nu}$ we obtain terms like
\begin{align}
\LL^{(2)}\supset& \bar{F}^a{}_{\mu\nu}\delta F^a{}_{\alpha\beta}\bar{g}^{\mu\alpha}\delta g^{\nu\beta}\supset\dot{A}\partial_0\delta A_{ij}\delta g^{ij}
\end{align}
where we see a coupling between both tensor modes that involves time derivatives. The large freedom in the choice of the building blocks of the Lagrangian leads to a very rich phenomenology for GWs, as the aforementioned oscillation explored in \cite{GWoscillationsGF} or the GWs opacity studied in \cite{Caldwell:2018feo}. In any case, it becomes clear that all these models can still be probed by using GWs astronomy without modifying $\cgw$.

\subsection{Temporally extended triad configuration}

The configurations considered in the two previous sections give rise to a background field configuration that respects some rotational invariance, either purely spatial or a combination of internal and spatial rotations. In this section we will consider a combination of these two configurations given by
\be
A^a_\mu=\phi^a\delta^0_\mu+A(t)\delta^a_\mu.
\label{eq:extendedtriad}
\ee
This configuration does not respect any rotational symmetry, even if the theory is provided with an internal $SO(3)$ symmetry. However, as shown in \cite{Jimenez:2016upj}, it is possible to restrict the interactions so that the energy momentum tensor becomes isotropic on-shell and, therefore, exact FLRW background solutions are still possible. This is a dramatically different mechanism to achieve homogeneous and isotropic solutions from those considered so far and, thus, it provides a new realisation of the Cosmological Principle.

Let us show the working mechanism with a very simple model that will serve as a proof-of-concept. The crucial property to guarantee that homogeneous and isotropic solutions exist for the configuration (\ref{eq:extendedtriad}) is that the Lagrangian only depends on the vector field without derivatives through $Y$ so that we will consider the Lagrangian $\LL=\mK(Y,Z_i)$, where $Z_i$ stands for the 11 possible Lorentz and $SO(3)$-scalars built out of the field strengths $F^a{}_{\mu\nu}$ (see for instance \cite{Piazza:2017bsd} for their explicit form, which is not relevant for our purposes here). Since $\phi^a(t)$ does not contribute to the field strengths, the sector containing the $Z_i$'s will automatically be isotropic and the only concern comes from the $Y$ sector. Variations of the corresponding action $\,S=\int\d^4x\sqrt{-g}\mK$ will thus take the form
\be
\delta \mS=\int\d^4 x\sqrt{-g}\Big[\mK_Y\delta Y+\mK_{Z_i}\delta Z_i-\frac12 \mK g_{\mu\nu}\delta g^{\mu\nu}\Big]\, .
\ee
The last two terms in this variation are identically isotropic and, therefore, will not be relevant for the following discussion. The only anisotropic contribution to the energy-momentum tensor in the FLRW metric with the configuration (\ref{eq:extendedtriad}) is then 
\be
T_{0i}\propto \mK_YA(t)\phi^a(t) \delta^a{}_i.
\ee
On the other hand, the equations for the $\phi^a$'s come from the same term and are given by
\be
\mK_Y\phi^a=0,
\ee
which clearly shows that the energy-momentum tensor is isotropic when the equations of the temporal components are satisfied. In fact, we can see the general property that $T_{0i}$ is proportional to the equation of motion of the temporal component in a homogeneous background (see also \cite{Jimenez:2009ai}). The branch with $\phi^a=0$ corresponds to the triad configuration of the previous section, while the branch with $\mK_Y=0$ is the genuine extended triad configuration, that we are considering in this subsection.

This configuration does not present new phenomenological consequences in the tensor sector with respect to the pure triad. The reason for this is that the $Z_i$-sector is exactly (off-shell) isotropic also in the extended triad configuration while the $Y$-sector does not contribute to the tensor modes. However, the isotropy violation of the extended triad introduces a preferred direction in the background that will reflect in the vector and scalar perturbations, giving a very distinctive signature of these models.  A full account of this phenomenology will be presented elsewhere, but the main feature can be easily understood by considering our proxy Lagrangian, whose quadratic form will read
\begin{align}
\LL^{(2)}\supset  \;&\mK_Y \delta^{(2)} Y +\mK_{Z_i} \delta^{(2)} Z_i\nonumber\\
&+\frac12\mK_{Z_i Z_j}\delta Z_i \delta Z_j+\frac12\mK_{Z_i Y}\delta Z_i \delta Y+\frac12\mK_{YY}(\delta Y)^2
\label{eq:L2extended}
\end{align}
where $\delta$ and $\delta^{(2)}$ stand for the first and second order perturbations respectively. 
 Again, the $Z_i$ sector exactly respects the background rotational invariance of the energy-momentum tensor and, consequently, there will not be any mixing from that sector. The above  expression also shows explicitly our previous statement that the $Y$ sector does not contribute to the tensor perturbations because $\delta^{(2)}Y$ enters multiplied by $\mK_Y$, that vanishes on shell, and $\delta Y$ does not receive contributions from the tensor modes. In fact, the first order perturbation of $Y$ is
\begin{align}
\delta Y=&\phi^2\delta g^{00}+A^2\delta^{ij}\delta g_{ij}+2A\phi^a\delta g^{0a}\nonumber\\
&+2\bar{g}^{00}\phi^a\delta A^a{}_0+2A\bar{g}^{ij}\delta A_{ij}\,,
\end{align}
where  $\phi^2=\vert\phi^a\vert^2$ and $\delta A_{ij}=\delta^a_i \delta A^a{}_j$. We then confirm that tensor modes do not contribute. Furthermore, we explicitly see the characteristic feature of this configuration that scalar and vector modes can mix via the background preferred direction provided by $\phi^a$ in the quadratic Lagrangian (\ref{eq:L2extended}) through the term  $\mK_{YY}(\delta Y)^2$.

\subsection{Gaugid configuration}
A particular class of models that allows for a different type of solutions is when the Lagrangian is provided with a set of abelian gauge symmetries, which was dubbed {\it gaugid} in \cite{Piazza:2017bsd}. For the multi-Proca interactions this can be easily achieved by setting all mass terms to zero, i.e., the action is only built out of the field strengths\footnote{There can be another alternative where the set of abelian gauge symmetries are non-linearly realised. If the gauge symmetry only involves up to first derivatives of the gauge parameter, the symmetry is the usual $U(1)$ up to field redefinitions \cite{Wald:1986bj}.} $F^a_{\mu\nu}$. For the non-abelian Yang-Mills theories with an $SU(2)$ symmetry, the same can be achieved by sending the gauge coupling constant $g\to0$ so that the full $SU(2)$ factorises into three $U(1)$'s, i.e., we have $\Fm^a_{\mu\nu}\rightarrow F^a_{\mu\nu}$. In both cases, the Lagrangian will be a function of the Lorentz- and $SO(3)$-scalars built out of the field strengths $F^a_{\mu\nu}$ mentioned above and whose explicit form can be found in \cite{Piazza:2017bsd}.

These models allow for solutions with the triad configuration that will lead to an {\it electric gaugid}  since, in that configuration, we have $F^a_{0i}=\dot{A} \delta^a_i$ and $F^a_{ij}=0$, so only the electric part of the fields contribute. The phenomenology for these models will differ from the one considered above in that the temporal components do not play any role because they become Lagrange multipliers instead of auxiliary fields. This will in turn result in fewer scalar modes for the perturbations.

So far, we have only considered homogeneous background configurations for the fields so that the required translational symmetry of the FLRW solutions is trivially realised. However, the gaugid models  make it possible to achieve good cosmological solutions with inhomogeneous fields, provided the three $U(1)$ gauge symmetries compensate for the inhomogeneities in the field configuration. This is analogous to what happens in solid inflation\footnote{Needless to say that the set-up of solid inflation could also be used to construct {\it solid dark energy} models.} \cite{Endlich:2012pz} (see also \cite{Gruzinov:2004ty}), which is based on three scalar fields $\phi^a$ with the configuration $\phi^a\propto x^a$ and where an internal shift symmetry for each scalar field makes up for the inhomogeneous field configuration. A similar construction can be used to have cosmological solutions with inhomogeneous vector fields. This was recently pursued in the model of gaugid inflation considered in \cite{Piazza:2017bsd}. The considered inhomogeneous configuration for the gauge fields is of the form 
\be
A^a{}_\mu=\frac12 B \epsilon^a{}_{i\mu}x^i
\ee
with $B$ a constant. In this configuration we have $F^a_{0i}=0$ and $F^a_{ij}=B\epsilon^a{}_{ij}$ so that $B$ represents a constant magnetic field and the configuration is therefore called {\it magnetic gaugid}. The Lagrangian is required to enjoy an internal $SO(3)$ symmetry together with three $U(1)$ gauge symmetries. The symmetry breaking pattern then goes as\footnote{Technically, the magnetic gaugid configuration does not break time translations, so that its corresponding generator would not be broken. However, in the late time universe, the presence of other matter fields (radiation, baryons, dark matter,...) will break time translations and that is why we do not include it in the final symmetry group.} $ISO(3,1)\times SO(3)\times [U(1)]^3\to ISO(3)$ so that each broken spatial translation can be compensated by a $U(1)$ transformation (analogously to the shift symmetry employed in solid inflation) and the broken spatial rotations are compensated by an internal one. Although this model was considered as an inflationary model, it can of course be used as a dark energy model as well so that it is also possible to have {\it magnetic gaugid dark energy}. An interesting property of this configuration is that, as shown in \cite{Piazza:2017bsd}, it naturally gives rise to a Chern-Simons type of interaction of the form $\epsilon^{ijk}\partial_i t_{jm}h_{mk}$ between both tensor modes.

A very remarkable phenomenological feature of dark energy models based on scalar fields is that non-linear interactions can lead to the appearance of screening mechanisms that allow to decouple the cosmological evolution of the scalar from its behaviour on small scales. Of course, the same will apply for the models considered in this work. We will highlight a mechanism that can be well-motivated within models of interacting dark matter and could naturally inscribe within models with the gaugid configuration. Some of those models can help alleviating some of the claimed small problems of $\Lambda$CDM and they rely either on self-interactions of the DM particles or interactions mediated by some gauge boson \cite{DMselfinteractions}. This gauge boson can be associated to gaugid dark energy, thus giving rise to interactions in the dark sector with specific signatures. In that case, we can imagine the gaugid to couple to some conserved current $J_a^\mu$ carried by the dark matter particles, which are assumed to share the same charge. It is then natural to expect that the charge density of dark matter will be proportional to its energy density. If we take a proxy model for the gaugid sector with Lagrangian $\LL=\mK(Z)$ and including an interaction through a coupling to the conserved current as $\LL_{\rm int}\propto A^a_\mu J^\mu_a$, the field equations around a static and spherically symmetric source with $J^a_\mu=\rho^a(r)\delta^0_\mu$ will be
\be
\vec{\nabla}\cdot(\mK _Z \vec{E}^a)=\alpha \rho^a\,
\ee
where $\alpha$ measures the strength of the coupling. As usual, we can integrate the above equation around a spherical shell comprising the source to obtain
\be
\vert \mK_ Z \vec{E}^a\vert=\frac{\alpha q^a}{4\pi r^2}
\ee
with $q^a=\int\rho^a\d^3x$ the total charge. Given our assumptions above, the total charge is expected to be proportional to the total mass of dark matter inside the spherical shell. We find then the expected screened solution when higher order interactions are included in the gaugid sector. Let us illustrate it with the simple Lagrangian $\mK=-1/4 Z(1+\frac{1}{\Lambda^4}Z)$ where $\Lambda$ is some scale controlling the non-linearities. At large distances we have $Z/\Lambda^4\ll1$ and we recover the usual Coulombian potential behaviour $ \vert \vec{E}^a\vert \propto r^{-2}$. However, below the non-linear scale determined by $r_{\rm NL}=\Lambda\sqrt{\vert\frac{\alpha q}{4\pi}\vert}$, the higher order term will take over and we have the screened solution $\vert \vec{E}^a\vert\propto r^{-2/3}$. This opens up the possibility for having different phenomenological signatures on small scales (inside dark matter haloes) and large scales and establishes a natural framework for new dark matter-dark energy interacting models.

\subsection{Approximate isotropic solutions: Oscillating fields}

In the precedent sections, we have focused on cosmological solutions where the required homogeneity and isotropy of the background solutions are exactly realised, i.e., the background field configurations exactly realise a residual $ISO(3)$ symmetry, which could happen to be realised only on-shell as in the extended triad configuration. However, other possibilities also exist where the background field configurations do not exactly respect those symmetries but deviations are sufficiently small as to be compatible with observations. We will be concerned here with the isotropy of the background configuration\footnote{Inhomogeneities are anyways considered in the perturbations and they must account for structure formation.} for which the CMB sets a constraint $\sim 10^{-3}$ for a dipole-like deviation and $\sim 10^{-5}$ for a quadrupolar deviation. This applies to models leading to Bianchi I universes as, e.g.,  anisotropic dark energy \cite{AnDE} or models with some background preferred direction  as it could be magnetic fields \cite{Campanelli:2006vb}, moving dark energy \cite{Moving} or vector fields \cite{Koivisto:2008xf}. We want however to highlight another less explored possibility that relies on oscillating fields. It is clear that a homogeneous spacelike vector field introduces some preferred direction (that could even vary in time depending on the polarisation of the vector field). However, it was shown in \cite{Cembranos:2012kk} that, as long as the oscillations of the vector field are fast enough as compared to the expansion rate of the universe, the corresponding energy-momentum tensor averaged over several oscillations of the field becomes isotropic. The same result also applies to non-Abelian Yang-Mills theories \cite{Cembranos:2012ng}  (it was even shown for arbitrary spin in \cite{Cembranos:2013cba}). These results were used in \cite{Cembranos:2016ugq} to show that oscillating coherent light vector fields can be good candidates for dark matter, similarly to axion-like dark matter models. We will not elaborate much further here on this possibility but we simply want to point out that these oscillating configurations can also give rise to accelerating cosmologies. In particular, for a power law potential of the form $V\propto A^n$, having an equation of state close to $-1$ requires a very small value of the exponent $n$, so that we essentially have a cosmological constant. However, the richer interactions structure of generalised Proca Lagrangians calls for a dedicated analysis as to explore which (if any) terms can easily give accelerating cosmologies with oscillating fields. This would open new interesting signatures since the background oscillating fields will affect non-trivially the perturbations, for instance sourcing the gravitational waves, generating a non-trivial slip or mixing different helicity modes \cite{Cembranos:2016ugq}. It is worth saying that the very oscillations of the background fields make this scenario quite cumbersome already at the background level, and the study of the perturbations quickly becomes a very challenging task, both analytically and numerically.

\section{Discussion}

In this work we have argued that the very restrictive bound on the GWs propagation speed inferred from GW170817/GRB170817A still leaves room for a wide class of dark energy models with interesting phenomenological consequences for structure formation and GWs probes. Throughout this work we have considered theories with vector fields and surveyed different configurations. We have shown that a purely temporal background field permits a non-minimal derivative coupling (with no analogous in the dark energy models based on scalar fields) which does not affect the GWs propagation but gives a non-trivial contribution to the scalar and vector perturbations. We have also considered the case of a triad configuration featuring a second tensor mode that can oscillate into GWs, giving distinctive signatures. We have extended the triad configuration to include temporal components and shown that this configuration can still support exact FLRW solutions while the breaking of isotropy in the background fields configuration induces a coupling between scalar and vector perturbations. We have discussed the case of gaugid  configurations that can provide dark energy models supported by inhomogeneous fields configurations. Finally, we have also commented upon the possibility of having background field configurations that only induce small deviations from isotropy and, in particular, the interesting case of oscillating fields, whose averaged energy-momentum tensor is in fact isotropic.

The results presented in this work make it apparent that dark energy models can still give a rich and interesting phenomenology without being in conflict with $\cgw=1$. We have considered theories with vector fields as a proof-of concept, but our results are not limited to that case. As an example, given the duality between vector fields and 3-forms in 4 dimensions, analogous configurations to the ones considered here are possible for 3-form dark energy models, with potentially similar phenomenologies. In particular, one can consider models with non-abelian $p-$forms with analogous symmetry breaking patterns as the ones discussed here. Dark energy models with additional types of fields like scalars \cite{Heisenberg:2018acv} or massive gravity \cite{Heisenberg:2017qka} will of course share some of these properties.

Another main message of this work is that interesting dark energy models are still possible without having to necessarily resort to contrived higher derivative and non-minimal couplings as those of the Horndeski Lagrangians and its extensions. It is perhaps more fructuous to turn to allegedly simpler models, as the ones discussed here, that explore fundamentally different dark energy models based on different symmetry breaking patterns or, equivalently, inequivalent realisations of the Cosmological Principle. In this respect, the classification presented in \cite{Nicolis:2015sra} or the approach discussed in e.g. \cite{EFTfluids} can provide a useful guidance. Let us finish by stressing that the cosmological evolution within these possibilities are fundamentally different owed to the different residual gauge symmetries of the perturbations, which is important for instance to understand the presence and behaviour of adiabatic modes and/or consistency relations \cite{Weinberg:2003sw,Hinterbichler:2012nm,Finelli:2018upr}.

\section*{Acknowledgements}
It is a pleasure to thank useful discussions with G. Ballesteros, M. Bartelmann, J. A. R. Cembranos, J. M. Ezquiaga, A. L. Maroto, F. Piazza, S. Tsujikawa and M. Zumalac\'arregui. J.B.J. acknowledges the support of the Spanish MINECO €œCentro de Excelencia Severo Ochoa Programme under grant SEV-2016-0597 and the projects FIS2014-52837-P and FIS2016-78859-P (AEI/FEDER).  LH thanks financial support from Dr.~Max R\"ossler, the Walter Haefner Foundation and the ETH Zurich Foundation.

\end{document}